\documentclass[aps,prc,preprint,groupedaddress,showpacs]{revtex4}
\usepackage{graphicx}
\begin{document}
\title{$J/\psi$ production in a gluon plasma produced in 
Au-Au collisions at the Relativistic Heavy Ion Collider}
\author{Bin Zhang}
\author{Donald L. Johnson}
\affiliation{Department of Chemistry and Physics,
Arkansas State University,
P.O. Box 419,
State University,
Arkansas 72467-0419, USA}
\date{March 20, 2004}

\begin{abstract}
Centrality dependence of $J/\psi$ production in 
Au-Au collisions at $\sqrt{s_{NN}}=200$
GeV is studied using the Glauber model plus the kinetic formation model. 
Initial $J/\psi$ production and destruction by incoming nucleons
are described by the Glauber model. $J/\psi$-charm equilibration
in the gluon plasma is described by the kinetic formation model.
We explore the possibility of $J/\psi$ suppression suggested by
the recent PHENIX data. We show that the $J/\psi$ yield monotonically
decreases with increasing centrality in the kinetic formation model 
when the charm mass is smaller than a critical mass. 
The final $J/\psi$ yield
is between the value after the Glauber suppression and the 
dynamical equilibrium value. This underscores the importance
of both $J/\psi$ production from d-Au collisions, which is essential
in determining the Glauber suppression, and high statistics Au-Au
data, which can further constrain final state $J/\psi$-charm
equilibration.
\end{abstract}


\maketitle

\section{Introduction}

Relativistic heavy ion collisions have been used for the search of
the Quark-Gluon Plasma, a deconfined nuclear matter that is believed to
have existed at about one micro-second after the 
big-bang \cite{muller95a,bass99a,qm2002a}. The
existence and properties of the Quark-Gluon Plasma are reflected 
in the particle spectra produced in these collisions. One important
particle that is sensitive to the Quark-Gluon Plasma production 
is the $J/\psi$ particle. As color screening in the Quark-Gluon Plasma 
can dissociate the $J/\psi$ particle, $J/\psi$ suppression
relative to the yield from the superposition of nucleon-nucleon
collisions can be used as a signal of the production of the 
Quark-Gluon Plasma \cite{matsui86a}.

In realistic collisions, the $J/\psi$ particle can also be 
destructed by incoming nucleons in the colliding nuclei. 
This is shown clearly by the proton-nucleus collision 
data \cite{e772a,e866a,na50a} where hot and dense nuclear 
matter is not expected to be produced.
Collisions between the $J/\psi$ particle and 
comovers produced in heavy ion collisions can further modify 
the $J/\psi$ yield \cite{shuryak79a,xu86a,gavin97a,kharzeev97a,cassing97a,gale99a,geiss99a,kahana99a,sa99a,spieles99a,vogt99a,blaizot00a,capella00a,gale01a}. 
Hence detailed phenomenological models
are necessary for the interpretation of $J/\psi$ production
data. By comparing theoretical expectations and experimental
results, an abnormal $J/\psi$ suppression in Pb-Pb collisions
at the Super-Proton-Synchrotron (SPS) \cite{na50b} has been identified 
which indicates the production
of hot and dense nuclear matter at the SPS energies.

At the Relativistic Heavy Ion Collider (RHIC) energies, there
are on average more than one charm-anti-charm quark pair produced in
each event. This opens the possibility of producing the $J/\psi$
particle in the deconfined phase via the reaction
$c+\bar{c}\rightarrow J/\psi+g$ \cite{thews01a}.
Other possibilities involving $J/\psi$ equilibration at the
hadronization time have also been proposed 
\cite{braunmunzinger00a,gorenstein01a,grandchamp01a}. Many interesting
studies have been made on the phenomenological consequences
of different production mechanisms 
\cite{zhang00a,gorenstein02a,grandchamp02a,zhang02a,zhang02b,andronic03a,bratkovskaya03a,grandchamp03a}. It was demonstrated by
the kinetic formation model that at collider energies,
due to charm recombination, $J/\psi$ enhancement may be
the signal of Quark-Gluon Plasma production, in contrast
to the traditional $J/\psi$ suppression scenario. 
Recently, data on $J/\psi$ production at RHIC have been 
released \cite{phenix03a,phenix03b}. 
No significant $J/\psi$ enhancement was observed.

In this paper, we will explore the possibility of $J/\psi$
suppression in the framework of the kinetic formation model.
In particular, we make the following variations of the
original kinetic formation model.

\begin{enumerate}
\renewcommand{\labelenumi}{(\arabic{enumi})}
\item
Recent pertubative Quantum Chromodynamics (QCD) studies of 
the charmonium spectrum 
\cite{brambilla01a,eidemuller01a,recksiegel03a,recksiegel04a} 
give a charm mass 
$m_c^{\overline{\textrm{MS}}}(m_c^{\overline{\textrm{MS}}})=1.243$ GeV,
different potential models \cite{eichten94a} 
give $m_c=1.5$ GeV to $m_c=1.8$ GeV,
preturbative QCD studies of charm production 
\cite{gavai95a,altarelli88a}
in hadron collisions give a charm mass from $m_c=1.3$ GeV to
$m_c=1.5$ GeV, a recent Nambu-Jona-Lasinio model 
calculation \cite{grandchamp03a} gives $m_c=1.6-1.7$ GeV. 
We will vary the charm
quark mass between 1.3 GeV and 1.9 GeV to study the effect
of the threshold of $J/\psi$ production on $J/\psi$-charm equilibration. 

\item
$J/\psi$ break-up cross section by a gluon was calculated in perturbative
QCD \cite{peskin79a,bhanot79a}. 
The final state is a pair of charm and anti-charm
quarks and the cross section decreases with the center-of-mass
energy of the $J/\psi$-gluon system. As the energy increases, 
the break-up is accompanied by additional parton radiation.
We will model this effect by using a $J/\psi$ break-up
cross section that is a constant in energy and varies between
1 mb and 6 mb and compare this with the dipole cross section
results.

\item
The original kinetic formation model uses an averaged uniform
parton density in the transverse plane. This may underestimate
charm-equilibration. In the following, we will use the Glauber
model to study particle production and this leads to
a better description of the evolution by local temperatures.

\item
We make use of the recent discovery by the PHENIX collaboration
and various theoretical approaches 
\cite{phenix02a,dokshitzer01a,batsouli03a,djordjevic03a,djordjevic03b,zhang03a,djordjevic04a,kharzeev04a}
that there is no strong charm energy loss in relativistic nuclear
collisions. We also use the recent PHENIX $J/\psi$
data in pp collisions \cite{phenix03a} to normalize our calculations.

\end{enumerate}

In the next section, the Glauber model and the kinetic formation
model will be introduced. We will use the Glauber model
to describe the destruction of initial $J/\psi$'s by incoming
nucleons and use the kinetic formation model to describe
$J/\psi$ equilibration in the Quark-Gluon Plasma. The results
will be presented in section \ref{sec_res}. Then we will discuss the 
implications of our study.

\section{\label{sec_mod}The Glauber model and the kinetic formation model}

\begin{figure}
\includegraphics[angle=0,width=4.8in,height=4.0in]{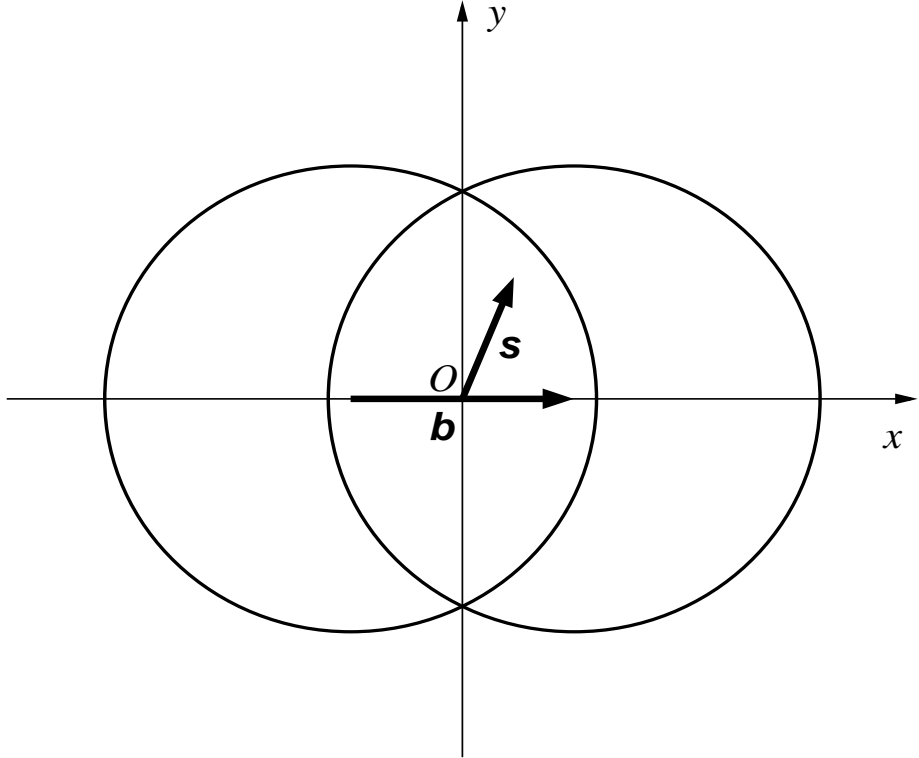}
\caption{\label{fig_coll_geo}Nucleus-nucleus collision transverse
geometry. The reaction plane is determined by the $x$-axis that goes
along the impact parameter vector $\vec{b}$ and the $z$-axis that
goes along the beam direction. In this figure, the $z$-axis points
out of the paper. The transverse position vector $\vec{s}$ starts
at the center of the overlapping region.}
\end{figure}

In nucleus-nucleus collisions, initial $J/\psi$'s are produced
by the collisions between incoming nucleons. Once a $J/\psi$ is produced,
other nucleons following the collision that produces the
$J/\psi$ will have opportunities to collide with it and turn it into
a charm-anti-charm pair. The number of $J/\psi$'s that
are produced by collisions between incoming nucleons and survive
further collisions with other nucleons can be written as
\begin{equation}\label{eq_AB2Jpsi}
N_{AB\rightarrow J/\psi}(b)=\int d\vec{s} \sigma_{NN\rightarrow J/\psi} 
t_A(\vec{s}+\frac{\vec{b}}{2}) 
t_B(\vec{s}-\frac{\vec{b}}{2}) 
S_A(\vec{s}+\frac{\vec{b}}{2})
S_B(\vec{s}-\frac{\vec{b}}{2}).
\end{equation}
In the above formula, $\vec{b}$ is the impact parameter vector,
and $\vec{s}$ is the transverse position. They are shown in 
Fig.~\ref{fig_coll_geo} along with the transverse coordinate system.
$\sigma_{NN\rightarrow J/\psi}$ is the $J/\psi$ production cross section
in nucleon-nucleon collisions. $t_A$ and $t_B$ are the nuclear
thickness functions of nucleus A and nucleus B, while $S_A$ and
$S_B$ are the average survival probabilities of a $J/\psi$
after passing through nucleons in nucleus A and nucleus B. The normalization 
by the probability of a nucleus-nucleus collision at impact 
parameter $b$, $\sigma_{AB}(b)$, is not included
as it only affects very peripheral events.

We will focus on $J/\psi$ production in Au-Au collisions. To simplify
the calculations, a nucleus is taken to be a hard sphere with uniform
nucleon density $\rho_N$ inside, and the radius is determined
by $R=r_0A^{1/3}$ ($r_0=1.12$ fm and $A$ is the number of nucleons
inside the nucleus). The number of survived $J/\psi$'s can now
be written as:
\begin{equation}\label{eq_AuAu2Jpsi}
N_{AuAu\rightarrow J/\psi}(b)=
\frac{\sigma_{NN\rightarrow J/\psi}}{\sigma_{J/\psi N}^2}
\int d\vec{s}
(1-\exp(-2\rho_N\sigma_{J/\psi N}z_0))
(1-\exp(-2\rho_N\sigma_{J/\psi N}z_0')),
\end{equation}
where $\sigma_{J/\psi N}$ is the $J/\psi$ break-up cross section
due to the collision with an incoming nucleon and $2z_0$ ($2z_0'$)
is the longitudinal length of the row of nucleons inside nucleus A 
(B) at transverse position $(x,y)$. More explicitly, 
$z_0=\sqrt{R^2-(x+\frac{b}{2})^2-y^2}$, 
and $z_0'=\sqrt{R^2-(x-\frac{b}{2})^2-y^2}$. The transverse
position integration $\int d\vec{s}$ becomes $\int dxdy$ over
the region $(x+\frac{b}{2})^2+y^2<R^2$ and $(x-\frac{b}{2})^2+y^2<R^2$.

After the passing of the two incoming nuclei, the $J/\psi$ is
immersed in the Quark-Gluon Plasma. As the initial gluon
number is much larger than the initial quark number at RHIC 
energies \cite{martin02a,eskola95a}, 
and in addition, light quark chemical equilibration
time is much longer than the expansion time \cite{wang97a}, 
the plasma will be approximated as a gluon plasma. 
As charm regeneration in the plasma is 
\cite{levai95a,lin95a,lin95b} neglegible, charm
quark number will be kept constant. There are indications that
charm quark thermal equilibration may be possible \cite{svetitsky88a}, 
and we will assume local thermal equilibrium.

$J/\psi$ equilibration in the plasma is
achieved through the reaction $J/\psi g\rightarrow c\bar{c}$ and its inverse
reaction $c\bar{c}\rightarrow J/\psi g$. Assume the rapidity distributions
are flat and focus on one fireball within one unit of rapidity,
the $J/\psi$ number $N_{J/\psi}$ satisfies
\begin{equation}\label{eq_rate}
\frac{dN_{J/\psi}}{d\tau}=\lambda_F N_c\rho_{\bar{c}}-
\lambda_D N_{J/\psi}\rho_g.
\end{equation}
In the above, $\tau$ is the proper time, $N_c$ is the number of charm
quarks per unit rapidity. $\rho_{\bar{c}}$ is the density of
anti-charm quarks and it equals the density of charm quarks,
$\rho_c$. $\rho_g$ is the density of gluons.

$\lambda_F$ is the reduced $J/\psi$ formation rate through the reaction
$c\bar{c}\rightarrow J/\psi g$. It equals 
$\langle \sigma_{c\bar{c}\rightarrow J/\psi g}v_{rel,c\bar{c}}\rangle$, 
where
$\langle\cdots\rangle$ denotes average over charm and anti-charm quarks,
and $v_{rel,c\bar{c}}$ is the relative speed between a charm quark
and an anti-charm quark. 
When the evolution is in thermal equilibrium, 
i.e., the evolution is controlled by one temperature $T$,
\begin{equation}
\lambda_F=\frac{1}{(2\alpha_c^2 K_2(\alpha_c))^2}\int_{z_0}^\infty
dz (z^2-4\alpha_c^2)z^2K_1(z)
\sigma_{c\bar{c}\rightarrow J/\psi g}(s=z^2T^2).
\end{equation}
We use $\alpha_c$ to stand for the mass of charm quark divided by
the temperature, i.e., $\alpha_c=m_c/T$. $z_0=\max(\alpha_c+
\alpha_{\bar{c}},\alpha_{J/\psi}+\alpha_g)$. $K_n(x)$ is the $n$th 
order modified Bessel function. $s$ is the center of mass energy squared.

Similarly, $\lambda_D$, the reduced $J/\psi$ destruction rate via
the reaction $J/\psi g\rightarrow c\bar{c}$ can be written as
\begin{equation}
\lambda_D=\frac{1}{8\alpha_{J/\psi}^2 K_2(\alpha_{J/\psi})}\int_{z_0}^\infty
dz (z^2-\alpha_{J/\psi}^2)^2K_1(z)\sigma_{J/\psi g\rightarrow c\bar{c}}
(s=z^2T^2).
\end{equation}

The detailed balance relation,
\begin{equation}\label{eq_det_bal}
\sigma_{c\bar{c}\rightarrow J/\psi g}(s)=\frac{d_{J/\psi g}}{d_{c\bar{c}}}
\frac{(k_{J/\psi g})^2}{(k_{c\bar{c}})^2}
\sigma_{J/\psi g\rightarrow c\bar{c}}(s),
\end{equation}
relates the two reduced rates. In Eq.~(\ref{eq_det_bal}), $d_{J/\psi g}=48$ is
the number of degrees of freedom for the $J/\psi g$ system, and
$d_{c\bar{c}}=36$ is the number of degrees of freedom of the 
$c\bar{c}$ system. $k_{J/\psi g}$ and $k_{c\bar{c}}$ are the center-of-mass
momenta of the $J/\psi g$ and $c\bar{c}$ systems. After using
the detailed balance relation, we can rewrite $\lambda_F$ as,
\begin{eqnarray}
\lambda_F&=&\frac{d_{J/\psi g}}{d_{c\bar{c}}}
\frac{1}{(2\alpha_c^2K_2(\alpha_c))^2}\int_{z_0}^\infty
dz(z^2-\alpha_{J/\psi}^2)^2K_1(z)
\sigma_{J/\psi g\rightarrow c\bar{c}}(s=z^2T^2) \nonumber \\
&=&\frac{d_{J/\psi g}}{d_{c\bar{c}}}
\frac{2\alpha_{J/\psi}^2K_2(\alpha_{J/\psi})}{\alpha_c^4K_2^2(\alpha_c)}
\lambda_D.
\end{eqnarray}

In Eq.~(\ref{eq_rate}), $N_c=N_{c,0}-N_{J/\psi}$ where $N_{c,0}$ is 
the number of charm quarks per unit rapidity at the initial time
$\tau_0$. $\rho_c=N_c/V(\tau)$. We use the Bjorken model \cite{bjorken83a}
for the evolution and the volume $V(\tau)=V_0\tau/\tau_0$. The gluon
density $\rho_g=\frac{g_g}{\pi^2}\zeta(3)T^3\approx 1.95 T^3=1.95
\frac{\tau_0T_0^3}{\tau}$.

\section{\label{sec_res}$J/\psi$ production at $\sqrt{s_{NN}}=200$ GeV}

\subsection{Initial conditions}

The initial gluon number, charm-anti-charm quark number, and $J/\psi$
number for central ($b=0$) Au-Au collisions will be specified. For
non-zero impact parameters, the total number of gluons and the 
total number of charm-anti-charm quarks follow binary collision
scaling, and the $J/\psi$ number follows 
Eq.~(\ref{eq_AB2Jpsi},\ref{eq_AuAu2Jpsi}).
The transverse density per unit rapidity of gluons (charm-anti-charm pairs)
at impact parameter $\vec{b}$ and position $\vec{s}$ is proportional
to $t_{Au}(\vec{s}+\frac{\vec{b}}{2})t_{Au}(\vec{s}-\frac{\vec{b}}{2})$
and is given by
\begin{equation}
\frac{dN_{g(c)}}{dyd\vec{s}}(\vec{b},\vec{s})=
4\rho_N^2z_0z_0'
\frac{1}{T_{AuAu}(b=0)}\frac{dN_{g(c)}}{dy}(b=0),
\end{equation} 
where $T_{AuAu}(b=0)$ is the nuclear thickness function at 
zero impact parameter. It equals
$\int d\vec{s}t_{Au}(\vec{s}+\frac{\vec{b}}{2})
t_{Au}(\vec{s}-\frac{\vec{b}}{2})=
4\rho_N^2\int d\vec{s}(R^2-s^2)=\frac{9}{8\pi}\frac{A^{4/3}}{r_0^2}$
with $A=197$ for Au nucleus.
The initial temperatures are local and calculated from
the initial gluon density distribution.
The $J/\psi$ transverse density per unit rapidity needs 
additional correction factor 
$S_{Au}(\vec{s}+\frac{\vec{b}}{2})
S_{Au}(\vec{s}-\frac{\vec{b}}{2})$.
It can be written as
\begin{equation}
\frac{dN_{J/\psi}}{dyd\vec{s}}(\vec{b},\vec{s})=
4\rho_N^2z_0z_0'
\frac{1}{T_{AuAu}(b=0)}\frac{dN_{J/\psi}}{dy}(b=0)
F(2\rho_N\sigma_{J/\psi N}z_0)F(2\rho_N\sigma_{J/\psi N}z_0'),
\end{equation}
where function $F(x)=\frac{1-\exp(-x)}{x}$.
The initial time is taken to be $\tau_0=0.2$ fm/c. 
In central collisions,
the initial gluon rapidity density is
$\frac{dN_g}{dy}=300$ and charm rapidity density is
$\frac{dN_c}{dy}=2.5$. They are in line with perturbative
Quantum Chromodynamics (QCD) based 
calculations \cite{wang91a,gyulassy94a,gavai95a}. The
initial $J/\psi$ rapidity density before correcting for the Glauber
suppression is $\frac{dN_{J/\psi}}{dy}=0.033$. This value comes from 
the superposition of $J/\psi$ production from nucleon-nucleon
collisions at $\sqrt{s_{NN}}=200$ GeV using the production
cross section measured by the PHENIX collaboration \cite{phenix03a}
and the nuclear
thickness function at zero impact parameter. The $J/\psi$ mass
is fixed at $m_{J/\psi}=3.1$ GeV. The charm quark mass $m_c$ will
be varied between $1.3$ GeV and $1.9$ GeV. The freeze-out temperature
$T_f$ is taken to be $0.15$ GeV.

\subsection{Cross sections}

The $J/\psi$-nucleon dissociation cross section $\sigma_{J/\psi N}$
will be varied among $0$, $4.4$ mb, and $7.1$ mb 
\cite{kharzeev97a,gavin97a,nagle99a,bennett99a,na50c}. 
The $J/\psi g\rightarrow
c\bar{c}$ cross section is taken to be a constant. We will vary
$\sigma_{J/\psi g\rightarrow c\bar{c}}$ to study 
$J/\psi$-charm equilibration. A comparison with the results
from a dipole cross section will be made to study the effect of
the energy dependence of the cross section.

\begin{figure}
\includegraphics[angle=0,width=4.8in,height=4.8in]{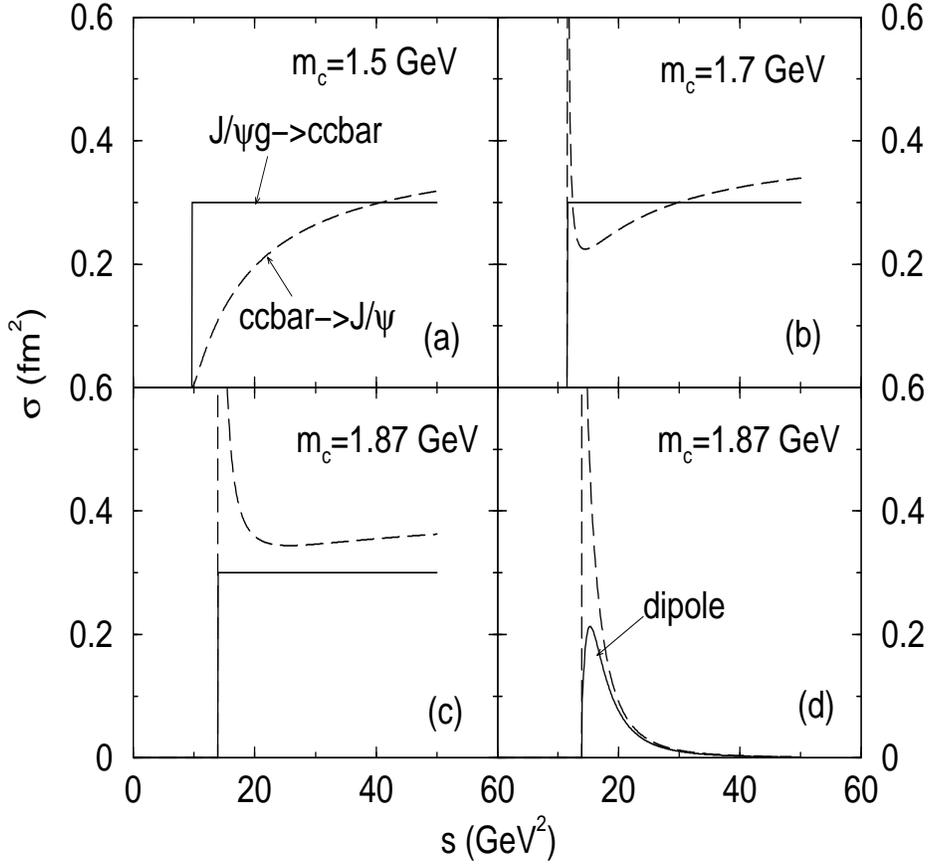}
\caption{\label{fig_xsecjpsi}The $J/\psi$ production and destruction
cross sections in a gluon plasma as functions of the center-of-mass
energy squared $s$. In (a)-(c), $\sigma_{J/\psi g}=3$ mb. In (d), 
$\sigma_{J/\psi g}$ is specified in Eq.~(\ref{eq_dipole}).}
\end{figure}

Fig.~(\ref{fig_xsecjpsi}) shows the $J/\psi$ production and
destruction cross sections as functions of the center-of-mass
energy squared $s$. In Fig.'s~(\ref{fig_xsecjpsi}a)-(\ref{fig_xsecjpsi}c), 
$\sigma_{J/\psi g\rightarrow c\bar{c}}$ is kept at a constant 
value of $3$ mb, which is about
the $J/\psi$ size in the vacuum. The inverse cross section,
$\sigma_{c\bar{c}\rightarrow J/\psi g}$, depends on the charm quark 
mass. When the charm quark mass $m_c=1.5$ GeV, which is smaller 
than the $J/\psi$ production threshold, the inverse cross section 
starts at zero and increases with $s$. It crosses the destruction 
cross section at $s\approx 40$ GeV$^2$. However, when $m_c=1.7$ GeV,
which is larger than the $J/\psi$ production threshold, there
is a peak at the threshold. When $s$ increases above the threshold,
$\sigma_{c\bar{c}\rightarrow J/\psi g}$ first decreases to below
$\sigma_{J/\psi g\rightarrow c\bar{c}}$. It then increases and crosses
$\sigma_{J/\psi g\rightarrow c\bar{c}}$ at $s\approx 30$ GeV.
The width of the peak at the threshold depends 
on the charm quark mass. The larger the mass, the wider the
peak. This can be seen by comparing $\sigma_{c\bar{c}\rightarrow J/\psi g}$
for $m_c=1.7$ GeV in Fig.~(\ref{fig_xsecjpsi}b) with 
$\sigma_{c\bar{c}\rightarrow J/\psi g}$ for $m_c=1.87$ GeV in 
Fig.~(\ref{fig_xsecjpsi}c). In addition, when $m_c$ increases, the crossing
position decreases. Eventually, $\sigma_{c\bar{c}\rightarrow J/\psi g}$
is always larger than $\sigma_{J/\psi g\rightarrow c\bar{c}}$ as
shown in Fig.~(\ref{fig_xsecjpsi}c). We also plot the $J/\psi g
\rightarrow c\bar{c}$ cross section based on a perturbative QCD 
calculation \cite{peskin79a,bhanot79a},

\begin{equation} \label{eq_dipole}
\sigma_{J/\psi g\rightarrow c\bar{c}}=\frac{2\pi}{3}\left(
\frac{32}{3}\right)^2\left(\frac{2\mu}{\epsilon_0}\right)^{1/2}
\frac{1}{4\mu^2}\frac{(k/\epsilon_0-1)^{3/2}}{(k/\epsilon_0)^5},
\end{equation}

and its inverse cross section in Fig.~(\ref{fig_xsecjpsi}d). 
In the above equation,
$\mu=m_c/2$ is the reduced charm quark mass in the
charm-anti-charm center-of-mass system. $k$ is the gluon 
momentum, and $\epsilon_0=2m_c-m_{J/\psi}$ is the binding 
energy of $J/\psi$. For $m_c$ equals the D meson
in vacuum mass $1.87$ GeV, the $J/\psi$ destruction cross
section has a finite peak near the threshold. The
peak of the cross section is about $2$ mb. The
production has an infinite peak at the threshold, and it is always 
larger than the destruction cross section.

\begin{figure}
\includegraphics[angle=0,width=4.8in,height=4.8in]{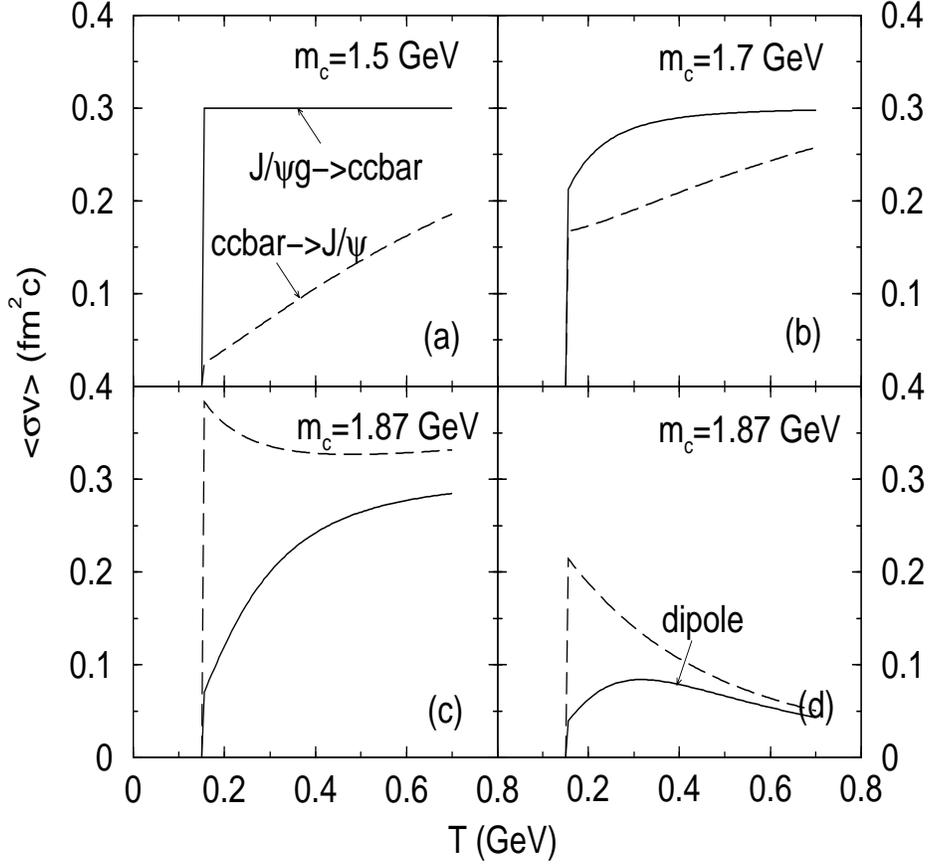}
\caption{\label{fig_sigmav_jpsi}The reduced reaction rates, 
$\langle\sigma_{J/\psi g\rightarrow c\bar{c}} v\rangle$ and
$\langle\sigma_{c\bar{c}\rightarrow J/\psi g} v\rangle$, 
as functions of the temperature of the plasma.}
\end{figure}

The reduced $J/\psi$ production rate 
$\langle\sigma_{c\bar{c}\rightarrow J/\psi g} v\rangle$ and
destruction rate
$\langle\sigma_{J/\psi g\rightarrow c\bar{c}} v\rangle$
for different cross section choices are
plotted in Fig.~(\ref{fig_sigmav_jpsi}). For $m_c$ smaller
than the $J/\psi$ production threshold, when
$\sigma_{J/\psi g\rightarrow c\bar{c}}$ is a constant, 
the reduce $J/\psi$ destruction rate is always a constant and
it equals $\sigma_{J/\psi g\rightarrow c\bar{c}}c$.
In Fig.'s~(\ref{fig_sigmav_jpsi}a)-(\ref{fig_sigmav_jpsi}c),
$\sigma_{J/\psi g\rightarrow c\bar{c}}=0.3$ fm$^2$.
For $m_c=1.5$ GeV, which is smaller than the $J/\psi$ production 
threshold, the destruction rate is a constant and the production
rate increases with the temperature. The destruction rate is always
larger than the production rate. 
As $m_c$ increases, the production rate increases. 
When $m_c$ is larger than the $J/\psi$ production threshold,
the destruction rate decreases with increasing $m_c$. 
For $m_c=1.7$ GeV, $\langle\sigma_{J/\psi g\rightarrow c\bar{c}}v\rangle$
starts from a value smaller than $\sigma_{J/\psi g\rightarrow c\bar{c}}c$.
It increases with temperature to approach 
$\sigma_{J/\psi g\rightarrow c\bar{c}}c=0.3$ fm$^2$c.
$\langle\sigma_{c\bar{c}\rightarrow J/\psi g} v\rangle$
has the same shape as the $m_c=1.5$ GeV case and increases
with the temperature of the plasma. It is smaller than the destruction 
rate up to $T=0.7$ GeV. When $m_c=1.87$ GeV, which is larger than the 
production threshold, the production rate is always larger
than the destruction rate. The production rate for $m_c=1.87$ GeV
decreases as the temperature of the plasma increases,
reflecting the decrease in the production cross section away
from the threshold. We also plot the rates for the case with the
dipole $J/\psi g\rightarrow c\bar{c}$ cross section determined by
Eq.~(\ref{eq_dipole}) in Fig.~(\ref{fig_sigmav_jpsi}).
$\langle\sigma_{c\bar{c}\rightarrow J/\psi g} v\rangle$ is still
larger than $\langle\sigma_{J/\psi g\rightarrow c\bar{c}} v\rangle$.
However, because the peak value of the dipole cross section 
is smaller than $0.3$ fm$^2$, also because the cross
section is exclusive and decreases with $s$, the rates
are smaller than the constant cross section case shown in
Fig.~(\ref{fig_sigmav_jpsi}c).

\subsection{Centrality and normalization}

We will use the number of participants to characterize the
centrality. The number of participants is calculated in the
Glauber model with hard sphere geometry according to 
\begin{eqnarray}
N_{part}(b)&=&\int d\vec{s} (t_A(\vec{s}+\frac{\vec{b}}{2})
(1-\exp(-\sigma_{NN}t_B(\vec{s}-\frac{\vec{b}}{2}))) \nonumber \\
&&+t_B(\vec{s}-\frac{\vec{b}}{2})
(1-\exp(-\sigma_{NN}t_A(\vec{s}+\frac{\vec{b}}{2})))) \nonumber \\
&=&8\rho_N\int_0^{R-\frac{b}{2}} dx \int_0^{\sqrt{R^2-(x+\frac{b}{2})^2}} dy
(z_0(1-\exp(-2\sigma_{NN}\rho_N z_0')) \nonumber \\
&&+z_0'(1-\exp(-2\sigma_{NN}\rho_N z_0))),
\end{eqnarray}
In the above, $\sigma_{NN}$ is the nucleon-nucleon inelastic cross 
section. We will use $\sigma_{NN}=40$ mb.

The $J/\psi$ central rapidity density is normalized to the
number of binary nucleon-nucleon collisions. We will plot
$B\frac{dN_{J/\psi}}{dy}|_{y=0}/N_{coll}$, where
$B=0.06$ is the $J/\psi$ to $e^+e^-$ or $\mu^+\mu^-$ decay
branching ratio, the number of binary nucleon-nucleon collisions 
is given by
\begin{eqnarray}
N_{coll}(b)&=&\sigma_{NN}\int d\vec{s} t_A(\vec{s}+\frac{\vec{b}}{2})
t_B(\vec{s}-\frac{\vec{b}}{2}) \nonumber \\
&=&4\rho_N^2\sigma_{NN}\int_0^{R-\frac{b}{2}} dx
\int_0^{\sqrt{R^2-(x+\frac{b}{2})^2}} dy z_0z_0'.
\end{eqnarray}

\begin{figure}
\includegraphics[angle=0,width=4.0in,height=4.0in]{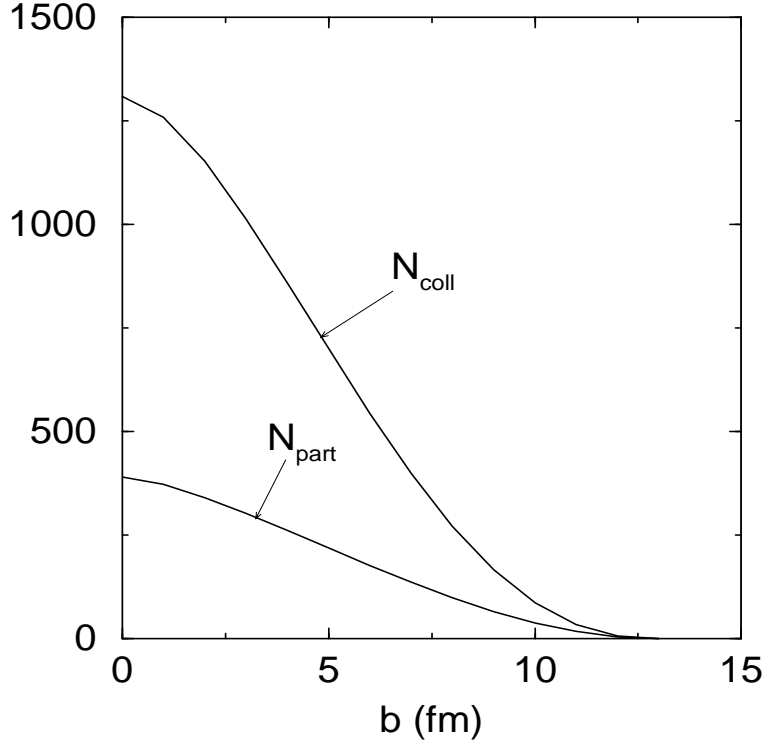}
\caption{\label{fig_nc_np}The number of participant nucleons $N_{part}$
and the number of binary nucleon-nucleon collisions $N_{coll}$ as
functions of the impact parameter.}
\end{figure}

The number of participant nucleons and the number of binary
nucleon-nucleon collisions are plotted in Fig.~(\ref{fig_nc_np}).

\subsection{Evolution of the gluon plasma}

\begin{figure}
\includegraphics[angle=0,width=4.0in,height=4.0in]{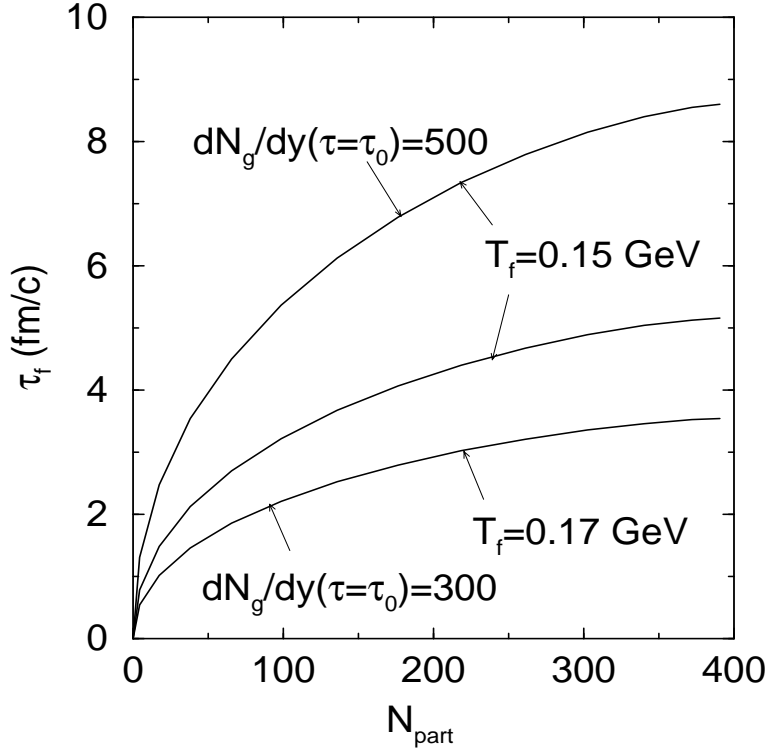}
\caption{\label{fig_tf_np}Freeze-out time $\tau_t$ as a function
of the number of participant nucleons $N_{part}$.}
\end{figure}

The evolution of the gluon plasma is described by the Bjorken 
model. We will use the latest freeze-out time $\tau_f$ to 
characterize the lifetime of the plasma. Fig.~(\ref{fig_tf_np})
shows that as the number of participant nucleons increases,
the lifetime of the plasma increases. The typical lifetime
for a plasma with an initial gluon rapidity density of 300
and a freeze-out temperature of 0.15 GeV is about 5 fm/c. 
Increasing the gluon rapidity density to 500 will bring
the lifetime to about 8.5 fm/c, while increasing the
freeze-out temperature to 0.17 GeV will decrease the
lifetime to about 3.5 fm/c.

\begin{figure}
\includegraphics[angle=0,width=4.0in,height=4.0in]{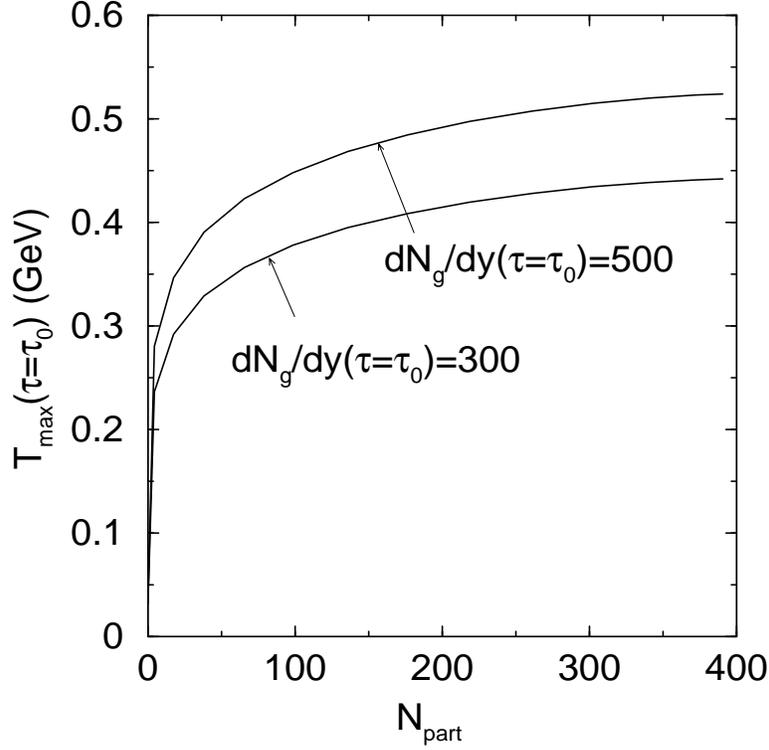}
\caption{\label{fig_temp_max_np}Initial temperature of the center cell
as a function of the number of participant nucleons $N_{part}$.}
\end{figure}

The temperature in the central transverse cell at the initial
time as a function of the number of participating nucleons
is shown in Fig.~(\ref{fig_temp_max_np}). It increases very rapidly
when the number of participating nucleons goes from 0 to 50
and changes only slowly when the number of participating nucleons
goes above 100. The typical initial temperature in the central
cell is about 500 MeV.

\subsection{$J/\psi$ production results}

$J/\psi$ enhancement was predicted by Thews \textit{et. al} using the
kinetic formation model. The PHENIX data excluded strong enhancement
of the $J/\psi$ particle. We are going to explore the
possibility of $J/\psi$ suppression from the kinetic formation model.
About $10$ $c\bar{c}$ pairs are produced on average in each
Au-Au event. We assume no strong initial state energy loss
of the produced $c\bar{c}$ pairs. The rapidity range of the
$c\bar{c}$ pairs is taken to be $\Delta y=4$. Now we vary the 
charm quark mass $m_c$, and observe the effect of changing 
the dynamical equilibrium of the $J/\psi$ particle through
the reactions $c\bar{c}\rightarrow J/\psi g$ and
$J/\psi g\rightarrow c\bar{c}$.

\begin{figure}
\includegraphics[angle=0,width=4.8in,height=4.8in]{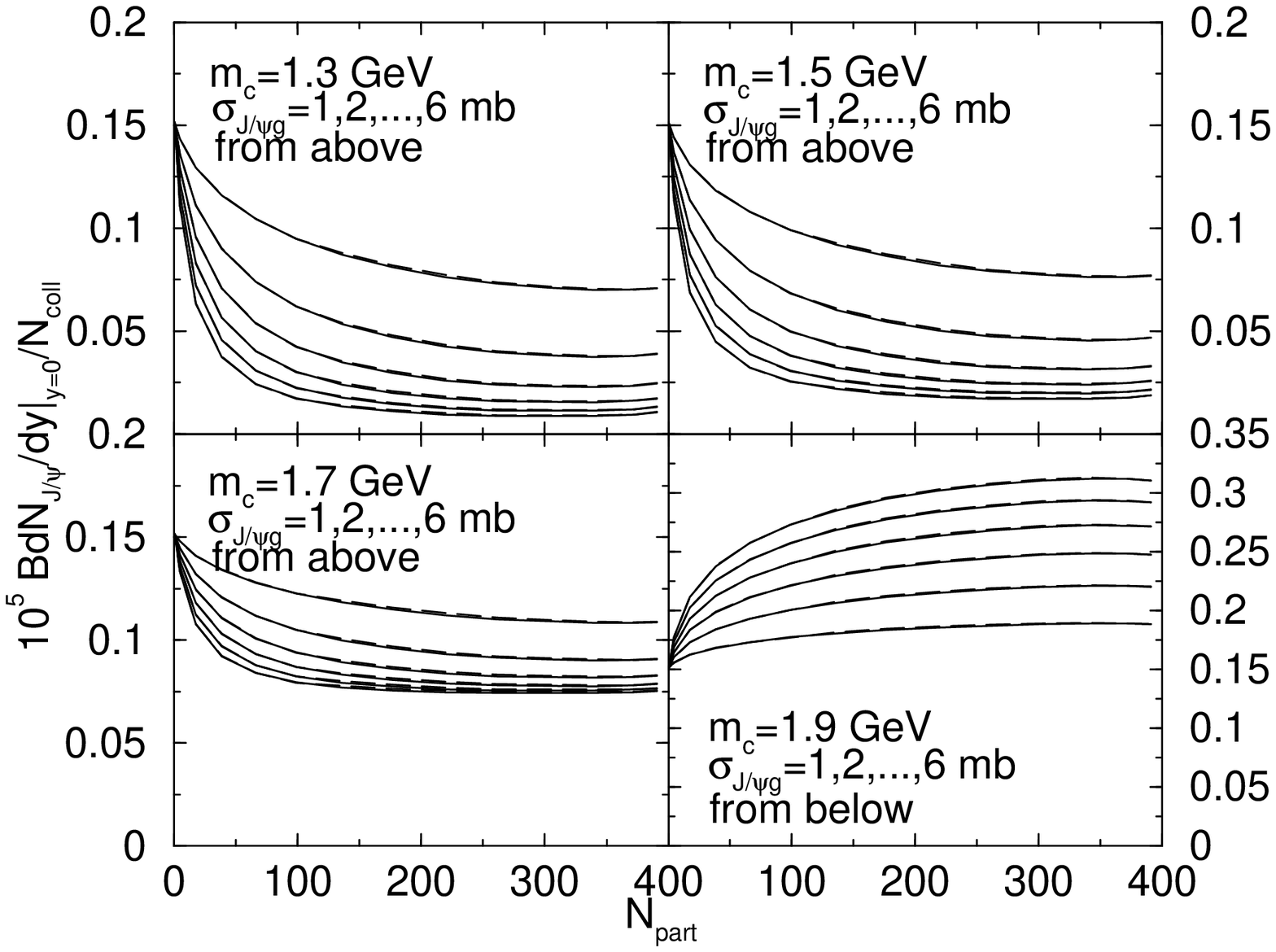}
\caption{\label{fig_bnj_nc_np7}Centrality dependence of 
$J/\psi$ central rapidity density per binary nucleon-nucleon collision
in Au-Au collisions when the suppression of $J/\psi$ by incoming
nucleons is negligible ($\sigma_{J/\psi N}=0$). Solid lines are for
$20\times 40$ cells and dashed lines are for $10\times 20$ cells.}
\end{figure}

The destruction of $J/\psi$ or its precursor (the correlated
$c\bar{c}$ pair that produces a $J/\psi$ when there are no
final state interactions) by collisions with incoming nucleons
 depends on the cross section 
$\sigma_{J/\psi N}$. When the colliding nuclei are highly Lorentz
contracted at high enough energies, the nuclei pass through each
other much earlier than the formation time of a $J/\psi$. 
The $J/\psi$ particle will not be strongly suppressed by
the incoming nucleons. We will use $\sigma_{J/\psi N}=0$ to
simulate this case. 
The centrality dependence of $J/\psi$ yields are shown 
in Fig.~(\ref{fig_bnj_nc_np7}).
When $m_c$=1.3, 1.5, 1.7 GeV, the final $J/\psi$ number is smaller
than the superposition of nucleon-nucleon collisions. The
larger the cross section, the smaller the final $J/\psi$ number.
When the $J/\psi$ destruction cross section $\sigma_{J/\psi g}$
goes up over about $5$ mb, the curves converge. This gives the
approximate dynamical equilibrium $J/\psi$ value. The
equilibrium value appears to be independent of centrality 
for mid-central and central Au-Au collisions. It
increases with the charm quark mass $m_c$. Also notice
that $m_c=1.7$ GeV is larger than the $J/\psi$ production
threshold. However, the initial $J/\psi$ is higher than
the equilibrium value, and the final $J/\psi$ number is smaller
than the initial. For $m_c=1.9$ GeV, The equilibrium value
is larger than the initial $J/\psi$ number
and the final $J/\psi$ number increases
with increasing $J/\psi$ gluon cross section. In the
$J/\psi$ suppression case, the final $J/\psi$ number decreases
with increasing centrality while in the $J/\psi$ enhancement case, it
increases with increasing centrality.

The adaptive step size Runge-Kutta method is used in solving 
the evolution equation. The transverse plane is
divided into $20\times 40$ cells ($20$ divisions in the $x$
direction from $-(R-\frac{b}{2})$ to $R-\frac{b}{2}$, and
$40$ divisions in the $y$ direction from $-\sqrt{R^2-(\frac{b}{2})^2}$
to $\sqrt{R^2-(\frac{b}{2})^2}$). Results are shown in 
Fig.~(\ref{fig_bnj_nc_np7}) as solid lines. Dashed lines are for 
the $10\times 20$ cell division case. They agree well with the $20\times 40$
cell division results.

At RHIC energies, the number of $J/\psi$ is much smaller than 
the number of charm-anti-charm quark pairs. One can approximate
the number of charm quarks at time $\tau$ by the initial number
of charm quarks, and solve Eq.~(\ref{eq_rate}) to obtain the 
solution
\begin{equation}
N_{J/\psi}(\tau_f)=\epsilon(\tau_f,\tau_0)\left[N_{J/\psi}(\tau_0)+
N_{c}^2(\tau_0)\int_{\tau_0}^{\tau_f}
\frac{\lambda_F(\tau)d\tau}{V(\tau)\epsilon(\tau,\tau_0)}
\right].
\end{equation}
Here,
\begin{equation}
\epsilon(\tau,\tau_0)=\exp\left\{
-\int_{\tau_0}^{\tau}\lambda_D(\tau')\rho_g(\tau')d\tau'
\right\}.
\end{equation}
The approximation solution gives no observable difference 
when the $J/\psi$ production rate is small, for example, 
when $m_c=1.3$ GeV. When the production rate is large,
for example, when $m_c=1.9$ GeV and $\sigma_{J/\psi g}=6$ mb, 
it can overestimate the $J/\psi$ yield (relative to
the exact solution) by about $4\%$.

\begin{figure}
\includegraphics[angle=0,width=4.8in,height=4.8in]{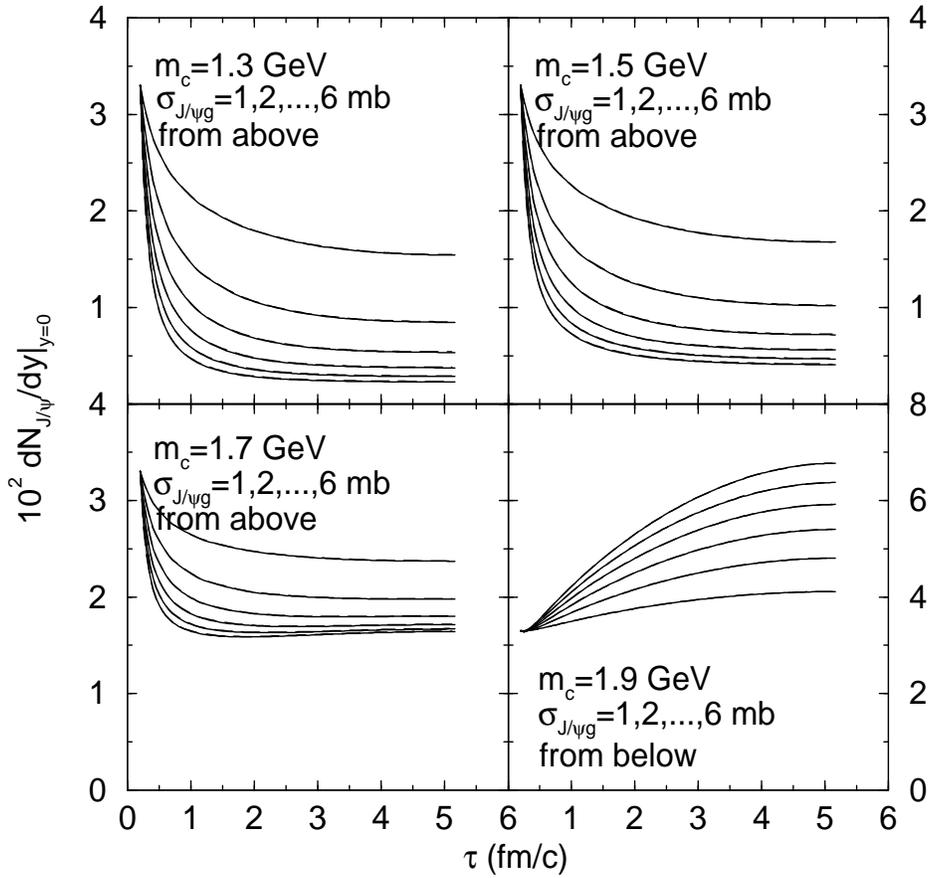}
\caption{\label{fig_xnjpsi_t_bb0.0}$J/\psi$ number evolution for
central ($b=0$) Au-Au collisions at $\sqrt{s_{NN}}=200$ GeV.}
\end{figure}

We show the time evolution of $J/\psi$ number in central
($b=0$) collisions in Fig.~(\ref{fig_xnjpsi_t_bb0.0}).
When $m_c=$1.3, 1.5, 1.7 GeV, the $J/\psi$ number changes
drastically before $\tau=$1 fm/c. When $m_c$=1.7 GeV and
$\sigma_{J/\psi g}=$5, 6 mb, the $J/\psi$ value reaches the lowest
point at about 1 fm/c, then it increases slightly. This is
because of the production of $J/\psi$ in central cells.
For $m_c=1.9$ GeV, the $J/\psi$ number changes over the 
entire evolution. The equilibration time is larger than
the system lifetime and results with different 
$\sigma_{J/\psi g}$ values do not converge and 
dynamical equilibrium of $J/\psi$ has not been reached
over the evolution.

\begin{figure}
\includegraphics[angle=0,width=4.8in,height=4.8in]{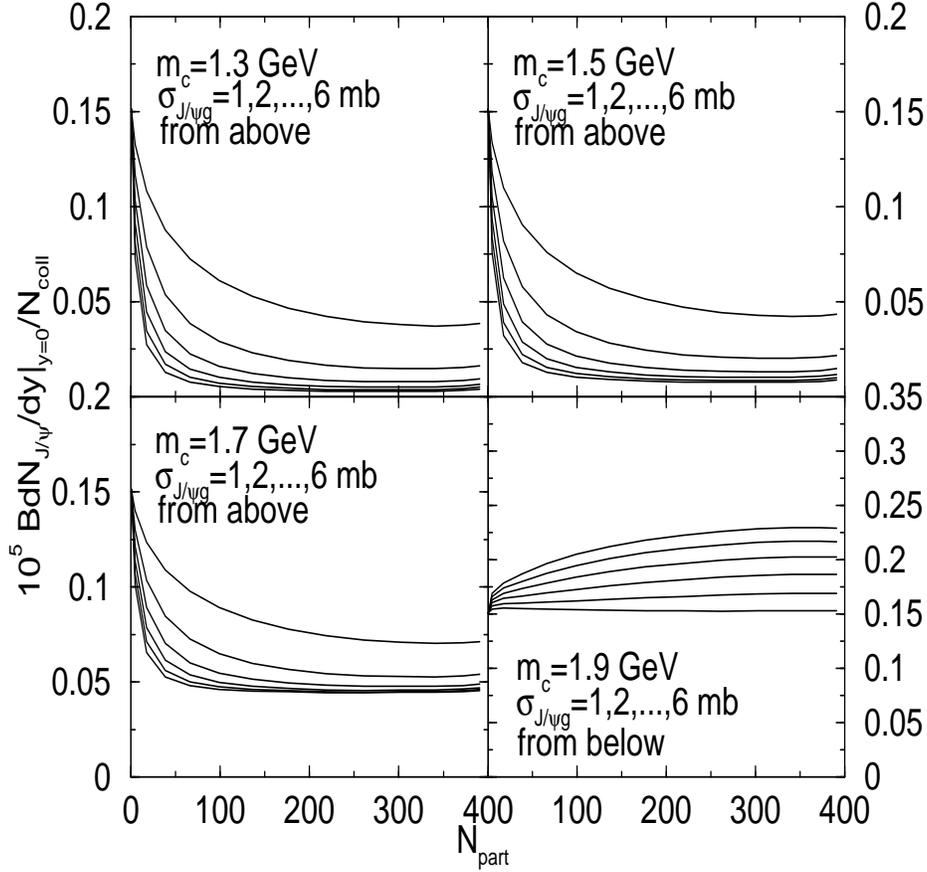}
\caption{\label{fig_bnj_nc_np7a}Centrality dependence of 
$J/\psi$ central rapidity density per binary nucleon-nucleon collision
when $\sigma_{J/\psi N}=0$ mb and $dN_g/dy(b=0)=500$.}
\end{figure}

\begin{figure}
\includegraphics[angle=0,width=4.8in,height=4.8in]{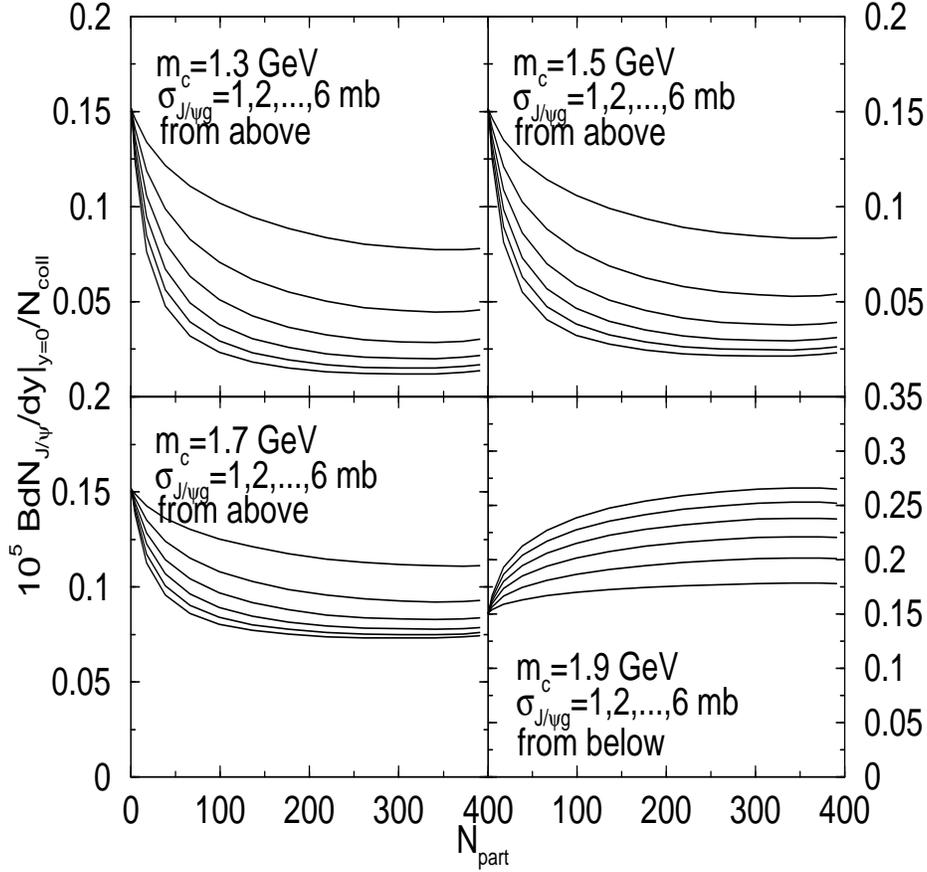}
\caption{\label{fig_bnj_nc_np7b}Centrality dependence of 
$J/\psi$ central rapidity density per binary nucleon-nucleon collision
when $\sigma_{J/\psi N}=0$ mb and $T_f=0.17$ GeV.}
\end{figure}

To check the sensitivity on parameters, we change the gluon rapidity
density from $300$ to $500$. Results are shown in 
Fig.~(\ref{fig_bnj_nc_np7a}).
As the initial gluon number increases, the initial temperature
increases, more $J/\psi$'s will be destructed during the evolution.
This leads to an overall decrease of the $J/\psi$ yield for all
charm mass values. If we keep the gluon rapidity density to
be $300$ and increase the freeze-out temperature from $0.15$ GeV
to $0.17$ GeV, no significant changes show up for $m_c=$1.3,1.5,
1.7 GeV as seen from Fig.~(\ref{fig_bnj_nc_np7b}). This indicates
that the equilibration time is much longer than the time
that goes from $T=0.17$ GeV to $T=0.15$ GeV. For the $m_c=$1.9 GeV
case, especially for large $\sigma_{J/\psi g}$ values, the final
$J/\psi$ values decrease significantly. For example, when
$\sigma_{J/\psi g}=6$ mb, the final $J/\psi$ value for central
collisions decreases from 0.31 to about 0.22. This shows
that for the $m_c=1.9$ GeV case, the final $J/\psi$ value is 
still far from equilibrium and the equilibration time is
not negligible compared with the time that goes from
$T=0.17$ GeV to $T=0.15$ GeV.

\begin{figure}
\includegraphics[angle=0,width=4.0in,height=4.0in]{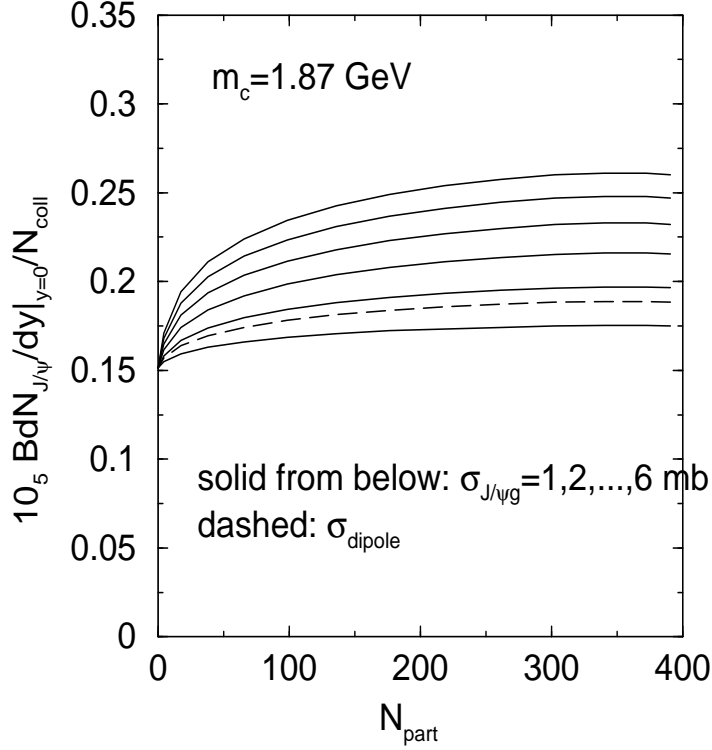}
\caption{\label{fig_bnj_nc_np6b}Centrality dependence of 
$J/\psi$ central rapidity density per binary nucleon-nucleon collision
when for $m_c=1.87$ GeV.}
\end{figure}

To see the difference between the constant $\sigma_{J/\psi g\rightarrow
c\bar{c}}$ case and the dipole form $\sigma_{J/\psi g\rightarrow
c\bar{c}}$ specified in Eq.~(\ref{eq_dipole}), 
we show the centrality dependence of $J/\psi$ production
in Fig.~(\ref{fig_bnj_nc_np6b}). 
The $J/\psi$ centrality dependence in the dipole
cross section case follows closely the constant cross section
case. It is between the two curves with $\sigma_{J/\psi g\rightarrow
c\bar{c}}=1.0$ mb and $\sigma_{J/\psi g\rightarrow
c\bar{c}}=2.0$ mb. The peak value of the dipole cross section
is about 2 mb. The $J/\psi$ yield is smaller than the constant
2 mb destruction cross section case because the dipole cross
section does not have destruction by high energy gluons with
additional radiation involved. The enhancement relative 
to the superposition of nucleon-nucleon production
when $\sigma_{J/\psi g\rightarrow c\bar{c}}=2.0$ mb
is about 1.2 times larger than the enhancement when the dipole
cross section is used, independent of centrality. 
In other words, the dipole cross
section case can be described by the constant cross section
case with an effective $K$ factor. We notice that the dipole
cross section and the constant cross section have very
different dependences on the center-of-mass energy 
(Fig.~\ref{fig_xsecjpsi}). The above comparison indicates
that the centrality dependence of $J/\psi$ production
reflects the overall production rate and can not 
differentiate between detailed cross section shapes.

\begin{figure}
\includegraphics[angle=0,width=4.8in,height=4.8in]{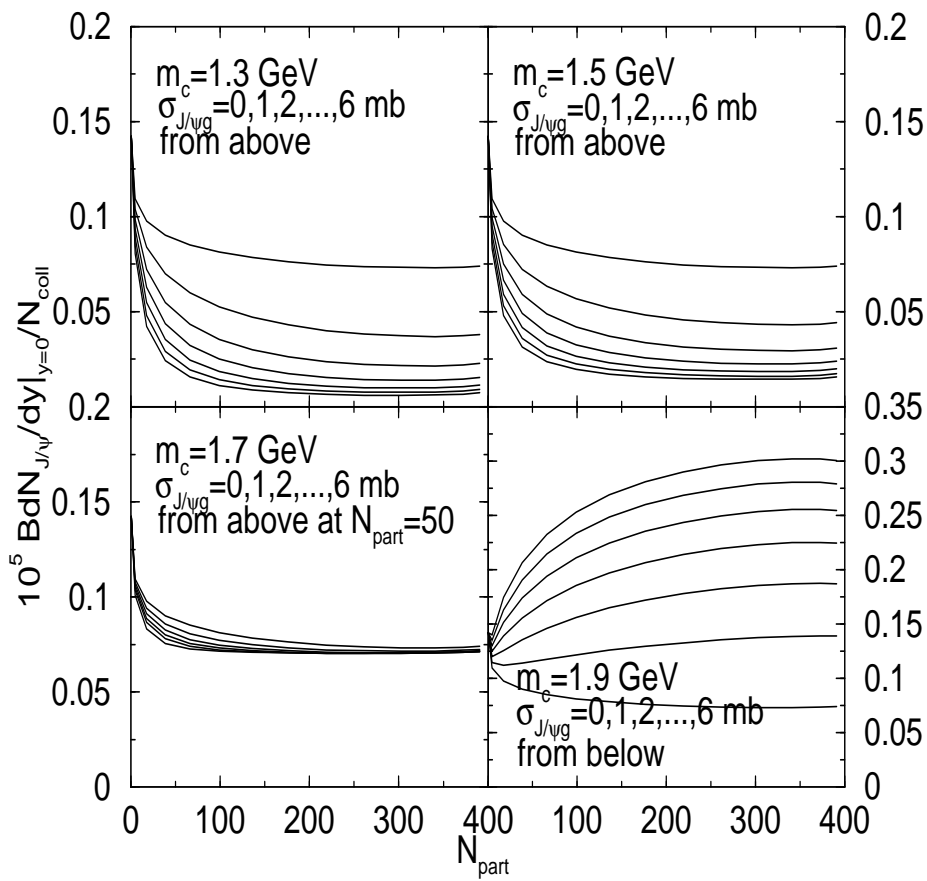}
\caption{\label{fig_bnj_nc_np9a}Centrality dependence of 
$J/\psi$ central rapidity density per binary nucleon-nucleon collision
when $\sigma_{J/\psi N}=4.4$ mb.}
\end{figure}

\begin{figure}
\includegraphics[angle=0,width=4.8in,height=4.8in]{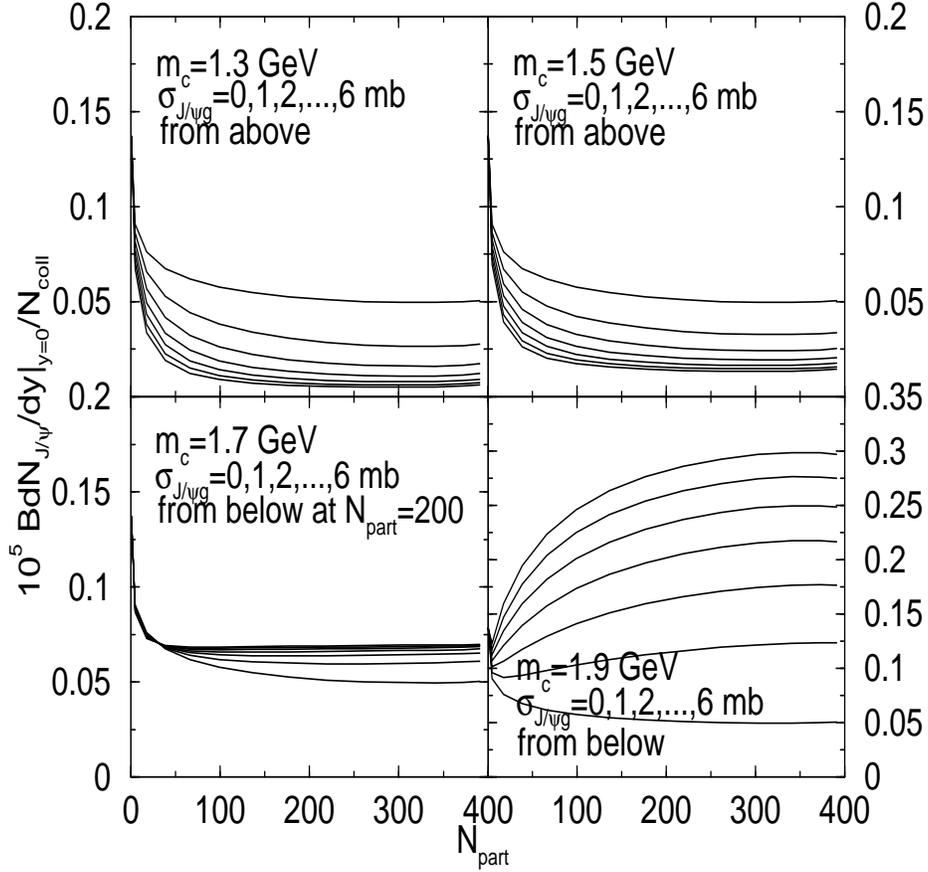}
\caption{\label{fig_bnj_nc_np9}Centrality dependence of 
$J/\psi$ central rapidity density per binary nucleon-nucleon collision
when $\sigma_{J/\psi N}=7.1$ mb.}
\end{figure}

Now we add in the Glauber suppression of $J/\psi$. We will first look
at the $\sigma_{J/\psi N}=4.4$ mb case. For $m_c=$1.3, 1.5, and 1.7 GeV, the
final $J/\psi$ values are still smaller than the initial values as
shown in Fig.~(\ref{fig_bnj_nc_np9a}). 
Because of the destruction by incoming nucleons, 
the final $J/\psi$ values and the initial are closer. For
$m_c=1.9$ GeV, the final $J/\psi$ values first decrease with
centrality and then increase with centrality. 
This centrality dependence differs from the monotonic 
centrality dependence when $m_c=1.3$, 1.5, 1.7 GeV. Because of 
suppression by the incoming nucleons, when $\sigma_{J/\psi g}$ is
small, the final $J/\psi$ values are no longer larger
than the superposition of $J/\psi$ from nucleon-nucleon collisions.
If $\sigma_{J/\psi N}$ is further increased to 7.1 mb, the initial
$J/\psi$ number after collisions with incoming nucleons is
even smaller as shown in Fig.~(\ref{fig_bnj_nc_np9}). 
The range between the initial and final $J/\psi$ values 
decreases for the $m_c=$1.3, 1.5 GeV cases.
For the $m_c=1.7$ GeV case, it goes below the
equilibrium value and the final state $J/\psi$ equilibration increases
the $J/\psi$ yield. When $m_c=1.9$ GeV, the range between
initial and equilibrium increases.

\section{\label{sec_sum}Summary and conclusions}

The $J/\psi$ production in Au-Au collisions at 
$\sqrt{s_{NN}}=200$ GeV is studied in the framework of the kinetic
formation model. It is shown that when the charm
quark mass $m_c$ is smaller than some critical value, 
the $J/\psi$ yield monotonically decreases with increasing centrality.
As $\sigma_{J/\psi g\rightarrow c\bar{c}}$ increases. 
the final $J/\psi$ yield moves away from the value
right after the Glauber suppression by the incoming 
nucleons and approaches the dynamical equilibrium value.
The $J/\psi$ production data from d-Au collisions
at the Relativistic Heavy Ion Collider will help
to fix the Glauber suppression and the higher
statistics Au-Au data will further help to understand
$J/\psi$-charm equilibration.

The current model does not have gluon nuclear shadowing. 
Further inclusion of shadowing will decrease the number of charm 
quarks produced and lower the number of $J/\psi$'s produced. 
Color screening \cite{matsui86a,datta03a,umeda03a,asakawa04a} 
will lead to more dissociation of $J/\psi$
and more suppression of final $J/\psi$ yield.
The inclusion of feeddown will lead to the same effect as higher
charmonium resonances are easier to be dissociated. Transverse
expansion will lead to earlier freeze-out and should not
change the qualitative feature of $J/\psi$ suppression. 
$J/\psi$-charm equilibration in the hadron phase is beyond
the scope of this paper. However, as the equilibrium
value in the hadron phase is smaller than the superposition
of production from nucleon-nucleon collisions, equilibration
in the hadron phase will also lead to $J/\psi$ 
suppression in Au-Au collisions.

\begin{acknowledgments}

We thank B.A. Li and Z. Lin for helpful discussions. We also
thank B. Hemphill and T. Franks for participation in the
early stage of the study. This work is supported
by the U.S. National Science Foundation under Grant No. PHY-0140046,
and the Arkansas Science and Technology Authority under 
Grant No. 01-B-20.

\end{acknowledgments}

\end{document}